\documentclass[prb,showpacs,twocolumn,preprintnumbers,amsmath,amssymb]{revtex4-1}
\usepackage{graphicx}
\usepackage{amssymb, amsmath, amsbsy}
\usepackage{color}
\usepackage{bm}

\begin{document}
\bibliographystyle{apsrev}



\title{
One-Dimensional Potential Model for Image States on 
Free-Standing Graphene.
}

\author{P.L. de Andres}

\author{P.M. Echenique}

\author{A. Rivacoba}

\affiliation{
Donostia International Physics Center,
Paseo Manuel de Lardizabal 4, 20018 Donostia, Spain.
}

\affiliation{
Materialen Fisika Saila, Kimika Fakultatea, UPV/EHU, 1072 P.K., 20080 Donostia, Spain
}

\affiliation{
Centro de Fisica de Materiales, CSIC-UPV/EHU, 20080 Donostia, Spain
}

\affiliation{
Instituto de Ciencia de Materiales de Madrid (CSIC),
Cantoblanco, 28049 Madrid, Spain.
}

\date{\today}

\begin{abstract}
In the framework of the non-local dielectric theory the static non-local
self-energy of an electron
near an ultra-thin polarizable layer 
has been calculated and applied
to study image-states near free-standing graphene.
The corresponding series of eigenvalues and eigenfunctions have been
obtained by solving numerically the one-dimensional 
Schr{\"o}dinger equation. We compare with 
the Rydgberg's series for a perfect metal and
with experimental values measured on graphene layers
grown on Ir and Ru surfaces.
For free standing films, the appearance of states with
binding energies in between the classical series is discussed.
\end{abstract}



\maketitle

Ultra thin stacks of graphene layers display a number of interesting properties
and potential applications
owing to the linear bands found near the 
{\bf K} point in the Brillouin zone.\cite{novoselov04,castroneto09} 
However, to extract its full potential, other
regions in the Brillouin Zone need to be considered;
in particular, unoccupied states in the vicinity of
${\bf \Gamma}$ have
also attracted attention due to their potential role
in transport of currents and heat.
Indeed, the dielectric response of very few layers of graphene 
is a key physical element to design
devices based in graphene.
The dielectric response is directly related,
and therefore can be investigated, by looking at
image states originated on the trapping of
external electrons in the region of unoccupied states between the 
Fermi level and the vacuum
level.\cite{echenique78}
Therefore, the experimental,\cite{hofer12,fauster12}
and theoretical\cite{silkin09}
study of image states
constitute an ideal probe to better understand the 
properties of ultra-thin graphene
layers.

The image force is a non-local effect asymptotically 
dominated by correlation effects.\cite{GMFF}
In order to study the infinite Rydberg series arising from the
image potential one needs to compute an effective one-dimensional
potential, V(z), representing the real part of the
quasi-static self-energy for an external unit probe charge. 
This self-induced potential is a continuos function spanning 
from inside the material, where it represents the exchange and
correlation energy, to the vacuum region, where
it should have the correct hydrogenic-like
asymptotic behavior, $-\frac{1}{4z}$.
Such a goal can only be obtained from a non-local spatial formalism,
since a local approach results in a correlation
potential decaying exponentially in the vacuum region,
following the density behavior outside the solid.\cite{langYkohn70}
For a self-consistent first-principles theory such a
non-local functional dependence can only be included
by means of costly numerical
calculations.\cite{alvarellos07} 
Therefore, it is useful and natural to search for simpler ways 
to obtain such an effective potential, which is
the basic ingredient needed to understand 
the physics of image states
bound by an ultra-thin polarizable layer like a few-layers
stack of graphene.
The simplest of these alternatives is to introduce a set of
fitting parameters to continuously join solutions valid
either inside or outside the solid.
This point of view has been taken, e.g., by Silkin et al. to study
image states in free-standing graphene,
joining a function with the correct asymptotic behavior 
to a local-density
approximation (LDA) calculation for the potential
inside an atom-thick graphene layer.\cite{silkin09}
This approach makes possible to
fit experimental data allowing its 
physical interpretation.\cite{hofer12b}
Its main weakness is its dependence on a few 
adjustable parameters, e.g. the choosing of 
a matching point in the vicinity of the surface,
which can influence results due to
the fast rate of change of the classical image potential
near its divergence at the origin (image plane), etc.

\begin{figure}
\includegraphics[clip,width=0.9\columnwidth]{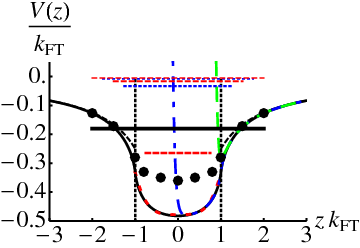}
\caption{(Color online).
Self-energy, $V(z)$,  
of a unit test charge at different positions
outside and inside a thin slab  
($k_{FT}=1$ and $d=1$).
Black continuos line: Eq. (1).
Black dots: RPA.
Black dashed line: asymptotic law with image
plane at $z_0=-\frac{1}{k_{FT}}$ ($\mid z \mid \ge d$).
Red dotted line: quartic approximation ($\mid z \mid \le d$) .
Long-dashed and dotted-dashed (blue and green): Eq. (1)
plus
a repulsive term, $R_{b}(z)= e^{-20 (z-b)}$.
Horizontal dashed and dotted lines 
show the first five eigenvalues (red and blue for
even and odd, respectively);
the horizontal thick line gives the approximate
value for the work function in graphene.
}
\label{fgr:LAM28}
\end{figure}

In this paper we analyze an alternative that is almost
parameter-free and makes a simple, flexible and accurate
basis for interpreting experimental results.
We use well-known models for the reflection of
electromagnetic waves at the surface (infinite barrier specular 
model,\cite{ritchie66}),
and for the 
non-local static dielectric
response (Fermi-Thomas and Random Phase Approximation\cite{pines})
so the desired self-energy can be obtained.\cite{deandres87}
In this approach only two free
parameters are neeed, i.e., the electronic medium polarizability which is
determined by the electron density of the material, and a geometrical one
give by the layer thickness.
This approach leads in a natural way to a potential
with proper physical features: it is continuous and finite over the
full spatial domain and it has the right asymptotic behavior towards the vacuum region.
Moreover, as for image states we are interested in regions in reciprocal space with
$\vec k$ near ${\bf \Gamma}$ and
energies between the vacuum level
and the Fermi energy,
it is well justified to model graphene as a polarizable 
electron gas with a quadratic energy dispersion, as seen from the 
relevant bands for 
graphene and graphite for these conditions.
The static dielectric response, $\epsilon(\vec k)$
has been modeled by a Random Phase
Approximation (RPA), and by its small $k$ expansion,
the Fermi-Thomas Approximation (FT).\cite{pines}
While the RPA yields a more accurate description of excitations
in the material, 
introducing the FT allows to write the potentials as analytical
expressions or quasi-analytical ones which merely depend 
on a final numerical step involving
the simple integration of a function decaying quickly
for large values of the argument.
The single parameter in these static models is
the screening constant, $k_{FT}$, 
that it is related to the density of
states at the Fermi level, $\frac{\partial n_0}{\partial \mu}$.
This value fixes the scale for energies, and its associated wavelength,
$\lambda_{FT}=\frac{2 \pi}{k_{FT}}$, the scale for lengths
(atomic units are used throughout the paper, except where
explicitly it is said otherwise). 
Taking graphite as a model 
(2 g/cm$^{3}$, 2s$^2$ 2p$^2$), typical values for graphene 
are $k_{FT} \approx 1$ ($r_s \approx 2.5$),
although its precise value may depend on factors like doping,
external potentials, etc; this is accommodated in our results through
the scaling with $k_{FT}$.
The other parameter needed to characterize a thin slab is its
width, $2d$.
For a single atom thick layer of graphene a reasonable value for $d$,
should be related to the spatial extension of $\pi$ carbon orbitals,
$d \approx 1$.

\begin{figure}
\includegraphics[clip,width=0.9\columnwidth]{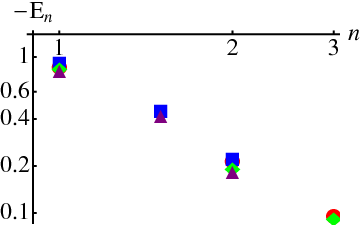}
\caption{(Color online)
Bound states energies, $-E_{n}$ (eV), for the
hydrogen-like series supported by the potential of the 
ultra-thin slab in Fig.~\ref{fgr:LAM28} (rectangles, blue).
These are compared with Whittaker's series (circles, red),
experimental values for Gr/Ru (triangles,
purple),\cite{hofer12}
and Gr/Ir\cite{fauster12} (diamonds, green).
The abscissas n labels are chosen to merely
follow the sequence of increasing energies.
}
\label{fgr:autoE}
\end{figure}

{\em Self-induced potential by an ultra-thin slab.}
For an external probe charge near a slab ($Q=1$) we seek the potential
acting on Q by the polarization charges induced in the medium
by Q itself.
This is obtained by computing the total potential,
and subtracting the charge's own naked potential.
To ensure the proper boundary conditions, and according to the
the specular reflection model at the surface, auxiliary
pseudo-media are introduced for the polarizable slab and the
vacuum that reduce the calculation to matching solutions obtained
in different regions of space for homogeneous media 
everywhere.\cite{GMFF}
Details for the thin slab, a vacuum gap between graphene
and a metal, and the metallic surface itself, along with
expressions, will be given in a forthcoming 
publication.
Within the FT approximation, this potential can be written
as an expression that only depends on a numerical 
integration ($\chi=\sqrt{\kappa^2+k_{FT}^2}$):

\begin{widetext}
\begin{equation}
V_{FT}(z > d)=-
\frac{k_{FT}^2}{2}\int_{0}^{\infty }
 \, d \kappa \,
\frac{e^{-2 \kappa z} }{ \left(
\chi+
\kappa \coth{
\left[
\chi d \right]}\right)
\left(\chi+\kappa
\tanh\left[\chi d \right]\right)}
\label{eq:FTvI}
\end{equation}
\noindent

\[
\label{eq:FTvII}
V_{FT}(0<z \le d)=
\int_{0}^{\infty } \, d \kappa \bigg \lbrace 
\frac{\chi+e^{4 \chi d} 
\left(\kappa e^{2 \chi z }+\kappa -\chi \right)}
{2 \left(
e^{4 \chi d}
-1\right) \chi}
+
\frac{\kappa}{2 \chi}
\bigg \lbrack
-
\frac{2 \kappa 
 \left(
(\chi+\kappa)
e^{2 (2 d+z)\chi }
+
(\chi-\kappa)
\right)}
{(2 \kappa^2+k_{FT}^2)
\left(e^{4 d \chi}-1\right)  +
2 \kappa \chi
\left(e^{4 d \chi}+1\right)}
\]

\[
+
\frac{e^{-2 \chi z}+1}{e^{4 \chi d}-1}
-
\frac{e^{-2 \chi (d+z) } 
\left(1+e^{2 \chi z}\right) \kappa 
\left(\kappa+\chi \left(e^{2 (d+z) \chi
}+\cosh \left[2 \chi d \right]\right) 
\text{csch} \left[2 \chi d \right]\right)}
{2 \kappa \chi
\cosh\left[2 \chi d \right]+
\left(2 \kappa^2+k_{FT}^2\right) \sinh \left[2 \chi d \right]}
\bigg \rbrack
\bigg \rbrace
\]

\end{widetext}

In Fig.~\ref{fgr:LAM28} we show the potential
for a slab occupying the region $-d \le z \le d$;
both in the FT approximation (black continuous line),
and in the RPA one (black dots).
In the all important region determining the Rydberg series
($\mid z \mid \ge d$), both approaches yield similar values
and agree with the correct asymptotic power-law.
Near the center of the slab, 
FT overestimates the interaction over RPA
by about 30-40\%, 
$$
\frac{V_{RPA}}{V_{FT}} \mid_{z=0} \approx 0.82-\frac{r_s}{12.5}
\quad ; \quad 2 \le r_s \le 6
$$
\noindent
a difference that reflects mainly in the first few states
with a significative weight at the center of the slab. 
In units of $k_{FT}$, we see that for $d>\frac{1}{k_{FT}}$,
$V_{FT}(d)=-1/3$,
and $V_{FT}(0)=-1/2$
(the Coulomb hole\cite{hedin99}).
Outside the slab,
for $z \ge \lambda_{FT}$,
the potential is well approximated by
a classical law,
$\frac{1}{4(z-z_0)}$,
corrected by 
an image plane, $z_0$. The value of
$z_0$ can be obtained by expanding the integrals for
$\kappa \approx 0$, giving the position of the
image plane in this model: $z_0=-\frac{1}{k_{FT}}$
(dashed line).
The potential inside the slab is in turn well
approximated by a quadratic or quartic fit 
(dashed red line Fig.~\ref{fgr:LAM28}).

\begin{figure}
\includegraphics[clip,width=0.45\columnwidth]{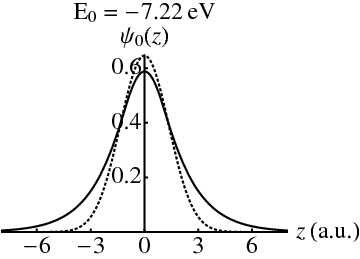}
\includegraphics[clip,width=0.45\columnwidth]{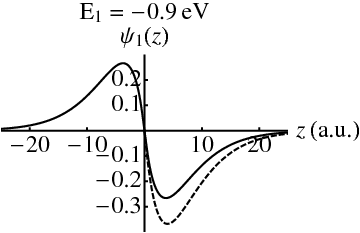} \\
\includegraphics[clip,width=0.45\columnwidth]{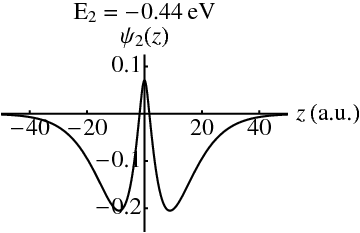}
\includegraphics[clip,width=0.45\columnwidth]{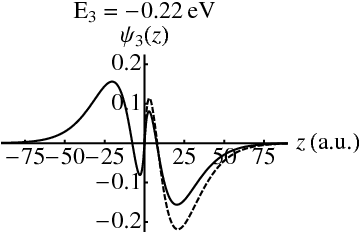}
\caption{
First four eigenfunctions for the potential displayed 
as a black continuous line in
Fig.~\ref{fgr:LAM28}. 
Eigenvalues are in eV and referred to the vacuum level.
For comparison, Whittaker's wave-functions ($1, 2$),
and the fitted harmonic oscillator wave function
($0$) are displayed (dashed). 
}
\label{fgr:autoF}
\end{figure}

{\em Eigenvalues and Eigenfunctions.}
We solve numerically the Schr\"odinger equation\cite{crandall82,shooting} to 
compute the eigenvalues and eigenfunctions 
corresponding to the Fermi-Thomas model
potential described above.
Quite generally, states can be 
labelled by the number of nodes ($n$),
with energies increasing as the number of nodes increases.
Furthermore, since V(z) in Eq.~1 is symmetric, eigenfunctions have
either even or odd parity, for even or odd $n$.
We show in 
Fig.~\ref{fgr:LAM28} 
the first five 
eigenvalues for V(z): dashed and dotted horizontal lines for
even and odd parities. 
It is worth noticing the structure of this series:
there is an isolated eigenvalue (the lowest one), while
the remaining states cluster near the vacuum level.
For standard densities (e.g., $1 < r_s < 10$) this first level
appears below $\lesssim -5$ eV; an estimate for the work function
in graphene (thick line).
Therefore, this $n=0$ eigenstate ($E_0=-7.2$ eV) does not fit in 
the standard definition for 
a Rydberg state, necessarily located between the vacuum
and Fermi levels to be observable in a standard
experiment. 
Furthermore, 
while wave-functions for Rydberg's states
are spatially located mostly in the vacuum region, 
$\psi_{0}$ is symmetric and peaks at the origin
(Fig.~\ref{fgr:autoF}, continuos
line in upper-left panel), being alike
to the ground state of a harmonic oscillator fitting
the bottom of the well 
($0.13 z^2$ a.u.,  $E'_0=-6.2$ eV),
but not to
states in the one-dimensional hydrogen-like series for a semi-infinite metal, that
go to zero at the image plane.

\begin{table}
\label{tbl:BE}
\begin{tabular}{r|rrrrrrrrrr}
\hline
l  &    &   1 &   & 2 &    & 3  &    &    &       \\
Gr/Ir\cite{fauster12} &    &  {\bf  -.83}&  & {\bf -.19} &  & {\bf -.09} &   &    &      \\ \hline
m+1  &    &   1  &   & 2 &    & 3  &    & 4  &    & 5 \\ 
$-\frac{1}{4z}$\cite{echenique78} &    & {\bf -.85}&   &{\bf -.21}&    &{\bf -.094}&    &{\bf -.053}&    &{\bf -.034} \\ \hline
n  &  0$^+$ & 1$^-$ & 2$^+$ & 3$^-$ & 4$^+$ & 5$^-$  & 6$^+$  & 7$^-$  & 8$^+$  & 9$^-$  \\ 
Eq.~(1)&-7.22&{\bf -.90}&{\bf -.44}&{\bf -.22}&-.15&{\bf -.096}&-.074&{\bf -.054}&-.044&{\bf -.034} \\ \hline 
m+1  &    &   1  &   & 2 &    &  3  &    & 4  &    & 5 \\ 
$R_{b=1}$ &     &-.61&    &-.17& & -.073 & & -.037 && -.032 \\ 
$R_{b=0}$ &     &-1.3&    &-.28& & -.13   & & -.083 && -.062 \\ \hline
l  &    & 1' & 1 & 2 &  &   &   &   &   &    \\ 
Gr/Ru\cite{hofer12} &    &{\bf -.80}&{\bf -.41}&{\bf -.18}&&&&&& \\ \hline 
l  & & 1$^+$  & 1$^-$ & 2$^+$ & 2$^-$ & 3$^+$ & 3$^-$  & &   &       \\ 
$z_0=3$\cite{silkin09} & &-1.47   &-.72&-.25&-.19&-.11&-.07&&& \\ 
$z_0=5$\cite{silkin09} & &-1.29   &-.57&-.24&-.17&-.11&-.06&&& \\ \hline
\end{tabular}
\caption{
Binding energies, $E_{n}$ (eV), compared for different cases.
Labels $n,m$ denote the number of nodes in wave-functions.
Data from other authors has been labelled as in the original papers.
We highlight in bold face numbers that can be compared across different
calculations or experiments and have been given an accompanying interpretation
in the text.
}
\end{table}

In  Table I and Fig.~\ref{fgr:autoE} we compare
eigenenergies calculated for the FT model with experimental values reported
for Gr/Ru,\cite{hofer12} 
Gr/Ir,\cite{fauster12} 
and with the limiting case of the
Rydberg series for a perfect metal
$E_{m+1}=-\frac{1}{32 (m+1)^2} ; \, m= 0, 1, 2, ...$.
where $m$ refers to the number of nodes for each state 
(notice that these wave-functions
only extend to $z>0$ half-space, and that the zero at the origin
is not counted as a node since it derives from the boundary conditions).
These eigenvalues can be conveniently obtained from multiple-scattering
techniques,\cite{echenique78} while wave functions are obtained
as the solution to Schr\"odinger's equation with $V_H(z)=-\frac{1}{4z}$ ($z>0$),
that can be reduced to Whittaker's 
differential equation.\cite{specialF}
The similarity of values found for the antisymmetric ($n^{-}$)
members of the series of states for V(z) and Rydberg's series is striking.
Such a similarity can be better understood by looking at the
corresponding wave-functions (Fig.~\ref{fgr:autoF}).
Boundary conditions make all Whittaker's wave-functions 
to go to zero at the origin, 
a condition that in the case of a symmetric well can
only be fulfilled by odd wave functions.
Moreover, if $n^{-}$ and $m$ give the number of nodes for odd wave-functions 
for the symmetric potential, and the Rydberg's one respectively, we
can make a one-to-one correspondence, $\frac{n^{-}-1}{2}=m$, 
that simply tells us that both sets of wave-functions have the same number
of nodes if $n^{-}$ is divided by two (only half-space) and the node
at the origin is discounted.
From a physical point of view, we can envisage two relevant limits:
a free standing slab producing a symmetric potential with states labeled
by the number of nodes and their parity, 
and a slab lying on a substrate where a particular surface gap may prevent penetration
of wave-functions inside the material leaving only half the space accessible for
image states.
Therefore, it is reasonable to argue that, according to our analysis,
for a free standing slab, or one interacting weakly with the substrate
support, it should be possible to observe the even states 
as new energies located in between the usual hydrogenic ones.
{\it It is interesting to notice that one of these states may have been
observed for Gr/Ru ($n=2$),\cite{hofer12}
while none of these have been reported for Gr/Ir}.\cite{fauster12}
This fact must be related to the strength of the interaction between
the supporting metal and the graphene layer, and
the penetration of wave-functions in both metals in a way
that goes beyond the scope of the current analysis.


\begin{table}
\label{tbl:xM}
\begin{tabular}{c|rrrrrrrrrr}
\hline
Whittaker's         & 0    &      & 1    &      & 2    &      & 3    &      & 4 \\ 
$\overline{z}$      & 3    &      & 12   &      & 29   &      &   51 &      & 79 \\ \hline
Eq.~(1) & 1    & 2    & 3    & 4    & 5    & 6    & 7    & 8    & 9  \\ 
$\overline{z}$      & 3    & 6    & 12   & 18   & 28   & 34   & 48   & 58   & 78 \\ \hline
\end{tabular}
\caption{
Expectation value 
$\overline{z}$ ({\AA})
for
Whittaker's and Eq.~(1) wavefunctions.
}
\end{table}

To assess how sensitive are the eigenvalues to the details
of the model potential we have added to $V(z)$ a repulsive term modeled 
as an exponential wall: $R_{b}(z)=e^{-20 (z-b)}$.
On a metallic surface such a "repulsive" term can originate because 
electronic gaps existing for particular surface orientations.
The resulting potential for the repulsive barrier located near the
slab surface ($b=1$, green dashed line in Fig. ~1) is similar
to the classical series with an image plane to prevent the
divergence at the origin ($\frac{1}{4(z+z_0)}$) and can be
understood by introducing a quantum defect in Rydberg's series. 
The barrier, on the other hand, can be introduced below the
surface (e.g., $b=0$, blue dashed-dotted line in Fig.~1),
and a numerical solution for Schr\"odinger's equation can be obtained. 
Results in Table I show that, for large m, eigenvalues tend to
the classical ones in accordance with the fact that the
average position of the electron is far away from the
surface, therefore mostly influenced by the
asymptotic behavior.
Table II gives the expectation mean values in {\AA},
$\overline{z}=\int_{0}^{\infty} \psi (z) \, z \, \psi(z) \, d \, z$,
for Whittaker's wavefunctions compared with the ones corresponding to
Eq.~(1).
These values compare well, which reflect the manifest similarity between wave functions
commented on Fig.~\ref{fgr:autoF}.
The fact that $\overline{z} \gg d$ for $n \gg 1$ determines a spatial region
where both FT and RPA dielectric functions lead to approximately the same potential,
suggesting that higher k-corrections to the dielectric function arising from
the random phase approximation are negligible, at least for $n \gg 1$ states.
Taking away the first level, largely affected by the details near the bottom
of the potential,
the rest of the series is only
modified by a percentage comparable
to differences found
in Table I between similar entries. 
Therefore, we argue that our assignation of levels is sound
and robust.

{\em Conclusions.}
Using standard models for the dielectric response and the reflection of electromagnetic
waves at a surface we have computed the static self-energy for an ultra-thin
slab mimicking a graphene layer. 
The self-induced potential goes continuously from the exchange and correlation energy
inside the material to the classical asymptotic image potential in the vacuum. 
Eigenvalues and eigenfunctions have been compared with Whittaker's classical series
and recent experiments on Gr/Ir and Gr/Ru. 
The odd members of the series for the slab show a remarkable resemblance to 
the solution of Schr\"odinger's equation for the
classical image potential (Whittaker's wavefunctions).
On the other hand, even wavefunctions arise as new 
states that differ from Whittaker's
in several key respects, e.g. their non-zero 
density probability at the origin.
For the case of films weakly interacting with a support 
some new states may consequently appear in between the classical ones, that
can be traced back to the even states in a free-standing
slab. Such a case could have been observed in recently measured experimental
values on Gr/Ru.

{\em Acknowledgments.}
This work has been financed by the Governments of Spain
(MAT2011-26534, and FIS2010-19609-C01-01),
and the Basque Country (IT-756-13).
Computing resources provided by the CTI-CSIC are 
gratefully acknowledged.



\end{document}